\input epsf.tex
\documentclass[12pt]{article}
\usepackage{graphicx}
\usepackage{psfig,picinpar,floatflt,amssymb,epsf}
\csname @addtoreset\endcsname{equation}{section}

\def\bseq{\begin{subequation}}  
\def\eseq{\end{subequation}}
\def\bsea{\begin{subeqnarray}}  
\def\esea{\end{subeqnarray}}

\def\Tilde#1{\widetilde{#1}}                    

\newcommand{\bbox}{\lower.2ex\hbox{$\Box$}}

\newcommand{\beq}{\begin{equation}}
\newcommand{\eeq}{\end{equation}}
\newcommand{\bea}{\begin{eqnarray}}
\newcommand{\eea}{\end{eqnarray}}
\newcommand{\ena}{\end{eqnarray}}

\renewcommand{\a}{\alpha}

\renewcommand{\d}{\delta}
\newcommand{\pa}{\partial}
\newcommand{\pab}{{\bar{\pa}}}
\newcommand{\g}{\gamma}

\renewcommand{\l}{\lambda}
\renewcommand{\L}{\Lambda}

\newcommand{\s}{\sigma}

\renewcommand{\t}{\tau}

\def\ch{\mbox{ch}}

\begin{document}

\begin{titlepage}
{\hbox to\hsize{December  2001 \hfill
{Bicocca--FT--01--32}}}
{\hbox to\hsize{${~}$ \hfill
{McGill--1--28}}}
\begin{center}
\vglue .06in
\vskip 25pt
{\Large\bf  An integrable noncommutative version of the sine--Gordon system}
\\[.45in]
Marcus T. Grisaru\footnote{grisaru@physics.mcgill.ca}\\
{\it Physics Department, McGill University \\
Montreal, QC Canada H3A 2T8 }
\\
[.2in]
Silvia Penati\footnote{silvia.penati@mib.infn.it}\\
{\it Dipartimento di Fisica, Universit\`a degli studi di
Milano-Bicocca\\ 
and INFN, Sezione di Milano, piazza della Scienza 3, I-20126 Milano, 
Italy}
\\[.5in]

{\bf ABSTRACT}\\[.0015in]
\end{center}
Using  the {\em bicomplex} approach we 
discuss a noncommutative system in two--dimensional
Euclidean space. It is described by an equation of motion 
which reduces 
to the ordinary sine--Gordon equation when the noncommutation parameter is removed,
plus a constraint equation which is nontrivial only in the noncommutative
case.
We show that the system has an infinite number of conserved 
currents and we give the general recursive relation for constructing them. 
For the particular cases of lower spin nontrivial currents 
we work out the explicit expressions and perform a direct check of 
their conservation. These currents reduce to the usual sine-Gordon currents in the commutative limit. We find classical  ``localized'' solutions to first 
order in the noncommutativity parameter and describe the Backlund 
transformations for our system. Finally, we comment on the relation of our noncommutative system to the commutative sine-Gordon system.

\vskip 30pt
${~~~}$ \newline

\end{titlepage}

\section{Introduction}

Field theories defined on noncommutative (NC) manifolds \cite{connes}
have been receiving 
 considerable attention in the last few years, primarily because of the
appearance of noncommutative geometries in strings \cite{string}
and matrix theory \cite{matrix}. The presence of spacetime noncommutativity 
has dramatic consequences on the dynamics of the fields and the quantum 
properties of the related theories (see \cite{review} for a review of the
subject and a quite complete list of references). In this context  it  is interesting 
 to investigate two-dimensional 
systems defined on a manifold where the two coordinates do not commute. 

Implementing noncommutativity on a two--dimensional Minkowski 
spacetime would necessarily involve the time coordinate in the
nonzero commutation relations.
However, it has been proven \cite{SST} that in general a 
noncommutation of the time variable affects the causality of the theory 
and its unitarity. Therefore, a well defined problem would be 
to consider  systems defined on a two--dimensional NC {\em euclidean}
space. It is well known that a selected class
of two-dimensional euclidean theories, i.e. conformal and integrable theories,
give a continuum description of two--dimensional statistical models at the
critical point \cite{BPZ} or perturbed away from the critical point along 
integrable directions \cite{zam}.         
In particular, one may be interested in the formulation of two--dimensional
{\em  integrable} theories in 
NC geometry and their possible connections with statistical mechanics. 

Some examples of NC equations which admit an infinite number
of conserved currents have been constructed in \cite{DMH1, DMH3}, by using 
a gauged bi--differential calculus. 
In this approach, which works in the ordinary commuting case \cite{DMH2}
and can be extended in NC geometries, the equations of motion of 
integrable systems are obtained as nontrivial consistency conditions for the 
existence of two flat covariant derivatives. As a consequence, by solving
an associated linear equation, one can establish the existence of an infinite
chain of conservation laws.
 
In this paper we use this procedure to construct a NC integrable system
whose equations of motion reduce to the ordinary sine--Gordon equation in
the commutative limit. Precisely, what emerges in the NC case is  a system 
of two
coupled equations; one of them contains a sine interaction term and can be
thought as a natural noncommutative analogue of ordinary sine--Gordon
equation, whereas the other one has the structure of a conservation equation
and can be seen as imposing an extra constraint on the system.
Only in the limit of commuting geometry the constraint becomes 
trivial and the second one reduces to the standard well--known equation.
At  first sight, the appearance of two equations seems quite unexpected and restrictive.
However, as  will be clear later on, this can be traced back to the fact
that the $SU(2)$ group which is the natural symmetry group of ordinary 
sine--Gordon, in the noncommutative case is not closed under $\ast$--product
and any noncommutative extension of the system must naturally rely on 
$U(2)$. This implies that the group valued fields which enter the bicomplex
construction take values in $U(2)$. Since  $U(2)$ contains a noncommutative $U(1)$
subgroup, they develop a nontrivial trace part which is responsible for the
appearance of an extra constraint equation \footnote{This is similar to what 
happens in the $U(1)$ WZNW model which, in the NC case, becomes
nontrivial \cite{WZNW,NOS} and does not have an immediate relation 
with its commutative counterpart (free scalar theory).}.   

We then give a general prescription to generate
conserved currents,  thus proving the classical integrability of the system. 
These currents reduce to those of the sine-Gordon system in the commutative limit.
To complete the classical analysis we also study  ``localized''
(pseudo--solitonic) solutions.
For the one--soliton solution we determine its distortion away from the
ordinary one due to the noncommutation of the  coordinates. 
This analysis is carried out perturbatively in the deformation parameter. 

In this work we have not succeeded in constructing an action from which the equations of motion emerge. Although this is an important issue, we feel that
integrability is an equally important feature of the sine-Gordon system. The alternative possibility, of starting with the usual sine-Gordon action and making it noncommutative is not a viable option in this respect.

The paper is organized as follows. In the next Section we summarize the 
general procedure based on the definition of a bicomplex. In
the third Section we construct the NC equations of motion of our system.
In Section 4 the iterative formula for constructing  an infinite number of
conserved currents is given and
the first two nontrivial currents are written explicitly. For the first,
the NC stress tensor, an explicit check of its conservation is performed
up to  second order in the noncommutation parameter $\theta$. For the
spin 3 current which in the ordinary case is a total derivative, triviality
is checked up to the first order in  $\theta$.
Section 5 is devoted to the study of one--``soliton'' solutions 
perturbatively in $\theta$, and a generalization of Backlund tranformations 
is given to generate $n$--``solitons'' solutions. Finally, Section 6 contains
our conclusions and an outlook on possible developments.
Two Appendices follow: The first one presents the detailed derivation of the
conserved currents, whereas in the second one we have collected the 
trigonometric $\ast$--calculus and all the identities required to perform 
perturbative $\theta$--expansions.

\section{Generalities on NC bi--differential calculus}

In this Section we summarize the general procedure to obtain integrable
equations in a NC geometry as given in \cite{DMH1}. The basic idea is to
write these equations as the nontrivial flatness conditions for two 
covariant derivatives suitably defined. 

Given a noncommutative two--dimensional space with euclidean signature and 
complex coordinates   
\beq
z = \frac{x^0 + i x^1}{\sqrt{2}} \qquad ; \qquad 
\bar{z} = \frac{x^0 - i x^1}{\sqrt{2}}
\eeq
noncommutativity is encoded in the relation 
\beq
[ z , \bar{z} ] = \theta
\eeq
where $\theta$ is a real parameter. The algebra ${\cal F}$ of smooth functions
on NC ${\cal R}^2$ is endowed with the product 
\beq
(f \ast g)(z,\bar{z}) = 
e ^{\frac{\theta}{2} (\pa_z \pab_{\bar{\xi}} - \pab_{\bar{z}} 
\pa_{\xi})} f(z,\bar{z}) g(\xi, \bar{\xi}) 
|_{\xi =z, \bar{\xi} = \bar{z}} ~=~ e ^{\frac{\theta}{2} P} fg
\eeq 
where
\beq
 P f g \equiv (\pa f \pab g - \pab f \pa g)
\label{P}
\eeq
\noindent
As basic ingredients we consider: 

\noindent
1) The $N_0$--graded linear space  
${\cal M} = \oplus_{r \geq 0} {\cal M}^r$, where 
${\cal M}^0 = {\cal F}$ or more generally a noncommutative algebra of 
functions in ${\cal F}$;

\noindent
2) Two linear maps $d,\d : {\cal M}^r \rightarrow {\cal M}^{r+1}$ 
satisfying $d^2 = \d^2 = \{ d , \d\}=0$. 

\noindent
The triple $({\cal M}, d,\d)$ is called a {\em bicomplex}. 
Given the two differential maps, we consider the associated linear equation
\beq
\d\chi = \l d \chi
\label{linear}
\eeq
where $\chi \in {\cal M}^s$ for a given ``spin''  $s$ and $\l$ is
a real parameter.
If a nontrivial solution exists, we can write
\beq
\chi = \sum_{l =0}^{\infty} \l^l \chi^{(l)}
\eeq
with $\chi^{(l)} \in {\cal M}^s$ satisfying 
\beq
\d \chi^{(0)} =0 \qquad ; \qquad \d \chi^{(l)} = d \chi^{(l-1)} \quad , 
\quad l > 0 
\label{iterative}
\eeq
Therefore we can construct a chain of $\d$--closed and $\d$--exact forms
in ${\cal M}^{s+1}$  
\beq
{\cal X}^{(l+1)} \equiv d \chi^{(l)} = \d \chi^{(l+1)} \qquad , \qquad l \geq 0
\eeq
We note that in order to obtain an actual chain of $\d$--closed and 
$\d$--exact forms it is necessary to require $\chi^{(0)}$ not to be 
$\d$--exact, i.e. the cohomology $H_{\d}^{(s)}$ must be nontrivial. 

In general, if the cohomology $H_{\d}^{(s+1)}$ is trivial we are guaranteed 
that an infinite chain of $\d$--closed and $\d$--exact forms exists. 
If the cohomology is not trivial, it is the solvability of the
linear problem which assures the possible existence of an infinite chain
of $\d$--closed and $\d$--exact forms.
 
When the two differential maps are defined in terms of the ordinary 
derivatives with respect to the two coordinates in ${\cal R}^2$ (see for
instance eq. (\ref{bicomplex}) below), the integrability conditions for 
the linear
equation are trivially satisfied ($\d^2 = d^2 = \{ d,  \d\} = 0$ by
definition). In this case the eqs.
(\ref{iterative}) have the appearance of an infinite number of conservation 
laws. However they are not the ones we are interested in since they are not
associated to any second order integrable equation.

Nontrivial integrable equations can be obtained by considering a {\em gauged}
bicomplex. We introduce two connections $A$ and $B$ and define
\beq
D_d = d + A \ast \qquad \qquad D_{\d} = \d + B \ast
\label{covariant}
\eeq
The flatness conditions $D_d^2 = D_{\d}^2 = \{ D_d , D_{\d} \} = 0$ imply
\bea
&& {\cal F}(A) \equiv d A + A \ast A = 0 \nonumber \\
&& {\cal F}(B) \equiv \d B + B \ast B = 0  \nonumber \\
&& {\cal G}(A,B) \equiv dB + \d A + A \ast B + B \ast A =0 
\label{EOMgen}
\eea
which, for a suitable choice of the bicomplex, give rise to non-trivial 
equations of motion. In the commutative \cite{DMH2} and noncommutative
cases \cite{DMH1, DMH3}, many known examples of 
integrable equations can be obtained from (\ref{EOMgen}). 

Again, one may consider the linear problem associated to (\ref{covariant})
\beq
{\cal D} \chi \equiv (D_{\d} - \l D_d) \chi = 0   
\eeq
The equations (\ref{EOMgen}) can then be seen as integrability conditions
for the linear equation, since
\beq
0 = {\cal D}^2 \chi = \left[{\cal F}(B) + \l^2 {\cal F}(A)
- \l {\cal G}(A,B) \right] \chi
\eeq 
If this equation has solutions  $\chi \in {\cal M}^s$ of the form
$\chi = \sum_l \l^l \chi^{(l)}$, we obtain an infinite chain of identities
\beq
D_{\d} \chi^{(0)} = 0 \qquad ; \qquad 
D_{\d} \chi^{(l)} = D_d \chi^{(l+1)} \quad , \quad l > 0 
\label{iterative2}
\eeq
which can be used to construct 
$D_\d$--closed and $D_\d$-exact forms ${\cal X}^{(l)}$, if $\chi^{(0)}$ 
is not cohomologically trivial.

As above, when the differential maps are defined in terms of ordinary 
derivatives, these equations can be interpreted  sometime as an infinite set of 
(nontrivial) conservation equations. 
However, in general, the  $\chi^{(l)}$ are nonlocal functions
of the coordinates (in the sense that they are defined in terms of integrals)
with no obvious physical interpretation.
As we will see in Section 4, conserved local objects can be constructed 
out of  the functions $\chi^{(l)}$ \cite{DMH2}.

\section{A NC sine--Gordon}

We apply the procedure described in the previous Section to
construct a noncommutative version of the sine--Gordon equation.

We consider the linear space ${\cal M} = {\cal M}^0 \otimes \L$, where 
${\cal M}^0$ is the space
of $2 \times 2$ matrices with entries in ${\cal F}$, and 
$\L = \otimes_{r=0}^2 \L^r$ is a two-dimensional graded vector space with the
 $\L^1$ basis 
$(\t,\s)$ satisfying $\t^2 = \s^2 = \t \s + \s \t =0$. 

For any matrix function $f \in {\cal M}^0$ we define two linear maps 
\beq
\d f = \pab f \t - R f \s \qquad ; \qquad
d f = - S f \t + \pa f \s
\label{bicomplex}
\eeq
where $R,S$ are constant matrices with $[R,S]=0$. It is easy to 
check that, as a consequence, the bicomplex conditions 
$\d^2 = d^2 = (d\d + \d d) =0$ are trivially satisfied.

To get nontrivial conditions, we introduce a {\em gauged} bicomplex 
by dressing the $d$ operator as
\beq
D   f \equiv G^{-1} \ast d (G \ast f) =
- L \ast f \t + ( \pa + M \ast ) f \s
\eeq
where $G$ is a generic invertible ($G \ast G^{-1} = G^{-1} \ast G =I$) 
matrix in ${\cal M}^0$ and
\beq
L = G^{-1} \ast S G \qquad ; \qquad M = G^{-1} \ast \pa G
\label{matrices}
\eeq
Now we require $({\cal M}, \d , D)$ to be a bicomplex. 
The condition $D^2=0$ implies 
$\pa L = [L, M]_{\ast}$ which one can check to be identically satisfied.
The last condition $\{D, \d\}=0$ gives instead the nontrivial equation
\beq
\pab M = [ R, L ]_{\ast}
\label{eq}
\eeq

In order to obtain a noncommutative version of the sine--Gordon equations
we choose the $U(2)$ group valued fields
\bea
&& R = S = \sqrt{\g} \left( \matrix { 0&0 \cr 
                            0&1 } \right) \nonumber \\
&& G = e_{\ast}^{\frac{i}{2} \s_2 \phi} = 
\left( \matrix{ \cos_{\ast}{\frac{\phi}{2}} &\sin_{\ast}{\frac{\phi}{2}} \cr
-\sin_{\ast}{\frac{\phi}{2}} & \cos_{\ast}{\frac{\phi}{2}} }\right)
\label{sgdif}
\eea
where $\ast$--functions are defined through their $\ast$--power series
(see Appendix B). As a consequence we have 
\bea
&& M = \frac{1}{2} \left( \matrix{ e_{\ast}^{\frac{i}{2} \phi}  \ast \pa 
e_{\ast}^{-\frac{i}{2} \phi} +  e_{\ast}^{-\frac{i}{2} \phi}  \ast \pa 
e_{\ast}^{\frac{i}{2} \phi}~&~  -i(e_{\ast}^{-\frac{i}{2} \phi}  \ast \pa 
e_{\ast}^{\frac{i}{2} \phi} -  e_{\ast}^{\frac{i}{2} \phi}  \ast \pa 
e_{\ast}^{-\frac{i}{2} \phi}) \cr
~~~& ~~~ \cr
i(e_{\ast}^{-\frac{i}{2} \phi}  \ast \pa 
e_{\ast}^{\frac{i}{2} \phi} -  e_{\ast}^{\frac{i}{2} \phi}  \ast \pa 
e_{\ast}^{-\frac{i}{2} \phi}) ~&~ e_{\ast}^{\frac{i}{2} \phi} \ast
\pa e_{\ast}^{-\frac{i}{2} \phi} + e_{\ast}^{-\frac{i}{2} \phi} \ast
\pa e_{\ast}^{\frac{i}{2} \phi} }\right)
\nonumber \\
&&~~~~~~~~~\nonumber \\
&&~~~~~~~~~\nonumber \\
&& L = \sqrt{\g} \left( \matrix{ \sin^2_{\ast}{\frac{\phi}{2}}
& -\sin_{\ast}{\frac{\phi}{2}} \ast \cos_{\ast}{\frac{\phi}{2}} \cr
-\cos_{\ast}{\frac{\phi}{2}} \ast \sin_{\ast}{\frac{\phi}{2}} &
\cos^2_{\ast}{\frac{\phi}{2}} }\right)
\eea
Computing $[R,L]$ we obtain
\beq
[R,L] = \g \left( \matrix{ 0
& \sin_{\ast}{\frac{\phi}{2}} \ast \cos_{\ast}{\frac{\phi}{2}} \cr
-\cos_{\ast}{\frac{\phi}{2}} \ast \sin_{\ast}{\frac{\phi}{2}} & 0 }\right)
\eeq
The equation (\ref{eq}) is a matrix equation in $U(2)$. In particular, 
the matrix $M$ has a nontrivial trace part, as a consequence of the
noncommutative nature of the $U(1)$ subgroup. Therefore,   
writing eq. (\ref{eq}) in components we obtain the two nontrivial 
equations for the field $\phi$
\bea
&& 
\pab \left( e_{\ast}^{\frac{i}{2} \phi}  \ast \pa e_{\ast}^{-\frac{i}{2} \phi} 
+ e_{\ast}^{-\frac{i}{2} \phi}  \ast \pa e_{\ast}^{\frac{i}{2} \phi}\right) 
~=~ 0
\nonumber \\
&&
\pab \left( e_{\ast}^{-\frac{i}{2} \phi}  \ast \pa e_{\ast}^{\frac{i}{2} \phi} 
- e_{\ast}^{\frac{i}{2} \phi}  \ast \pa e_{\ast}^{-\frac{i}{2} \phi}\right) 
~=~ i \g \sin_{\ast}{\phi}
\label{sg3}
\eea
where the identity (\ref{A1}) has been used. We note that in the limit
$\theta \to 0$ the first equation becomes trivial, whereas the second one 
reduces to the ordinary sine--Gordon equation
\beq
\pa \pab \phi = \g \sin{\phi} 
\eeq
In the noncommutative case, since 
$\pa e_{\ast}^{\phi} \neq e_{\ast}^{\phi} \ast \pa \phi$ (see eq.(\ref{A2})), 
both  equations
are meaningful and describe the dynamics of the field 
$\phi(z,\bar{z}, \theta)$.

\section{Conserved currents}

In the ordinary, commutative case, the derivation of the sine--Gordon 
equations from a bicomplex automatically guarantees \cite{DMH2} the 
existence of
an infinite chain of currents satisfying the conservation equations
\beq
\pab {\cal J}^{(l)} = \pa \Tilde{\cal J}^{(l)} \qquad , \quad l \geq 0
\label{conservation}
\eeq  
In this Section we extend those arguments to the noncommutative
case in order to prove the classical integrability of the system whose
dynamics is given by (\ref{sg3}). We will find a recursive procedure
to determine an infinite set of $\ast$--functions satisfying 
(\ref{conservation}). 

Quite generally, given a  $\ast$--invertible  function $\tilde{\chi} 
\in {\cal M}^0$ we define functions
\beq
J \equiv {\rm Tr}((\pa \tilde{\chi}) \ast \tilde{\chi}^{-1}) \qquad , 
\qquad \Tilde{J} \equiv 
{\rm Tr}((\pab \tilde{\chi}) \ast \tilde{\chi}^{-1})
\eeq
which satisfy the following identity
\beq
\pab J = \pa \Tilde{J} + [ \Tilde{J}, J]_{\ast}
\eeq
This is almost a conservation law, up to the commutator. To get rid of it, we
first observe that it can be written as \cite{DMH2}
\beq
[ \Tilde{J}, J]_{\ast} = \theta ( \pa \Tilde{J} \diamond \pab J - 
\pab \Tilde{J} \diamond \pa J )
\eeq
where we have introduced the new product 
\beq
f \diamond g \equiv \frac{\sinh{(\frac{\theta}{2}P)}}{\frac{\theta}{2}P} f g
\label{diamond}
\eeq
with the operator $P$ given in (\ref{P}). 
Therefore, if we introduce
\beq
{\cal J} \equiv J - \theta J \diamond \pa \Tilde{J} + \pa {\cal T}  
\qquad , \qquad \Tilde{\cal J} \equiv 
 \Tilde{J} -  \theta J \diamond \pab \Tilde{J} + \pab {\cal T}
\label{recipe}
\eeq
they satisfy the conservation equation
\beq
\pab {\cal J} = \pa \Tilde{\cal J} 
\label{conservation0}
\eeq
where ${\cal T}$ represents possible trivial terms. In particular, for  
an invertible  solution of   
\beq
\d \chi = \l D \chi
\label{SGlinear}
\eeq
defined as  a power series in $\l$, 
we can write for the functions ${\cal J}$ and $\Tilde{\cal J}$ 
\beq
{\cal J} = \sum_{l=0}^{\infty} \l^l {\cal J}^{(l)}
\qquad \qquad
\Tilde{\cal J} = \sum_{l=0}^{\infty} \l^l \Tilde{\cal J}^{(l)}
\eeq 
and from (\ref{conservation0}) we obtain an infinite set of conserved currents
associated to the equations of motion (\ref{EOMgen}).
We stress that it is the solvability of the linear equation (\ref{SGlinear})
which guarantees that the number of currents is infinite,
i.e. that the system is classically integrable.   

We now turn to the construction of the quantities ${\cal J}^{(l)}$ and
$\Tilde{\cal J}^{(l)}$. To simplify the notation we introduce the 
following functions
\bea
&& a \equiv \frac12 \left( e_{\ast}^{\frac{i}{2} \phi} \ast \pa 
e_{\ast}^{-\frac{i}{2} \phi} 
+ e_{\ast}^{-\frac{i}{2} \phi} \ast \pa e_{\ast}^{\frac{i}{2} \phi} \right) 
\nonumber \\
&& b \equiv \frac{i}{2} \left( e_{\ast}^{\frac{i}{2} \phi} \ast \pa 
e_{\ast}^{-\frac{i}{2} \phi} 
- e_{\ast}^{-\frac{i}{2} \phi} \ast \pa e_{\ast}^{\frac{i}{2} \phi} \right)
\eea
subject to the conditions  
\beq
\pab a = 0 \qquad; \qquad \pab b= \frac{\g}{2} \sin_{\ast}{\phi}
\label{eqmotion2}
\eeq
as follows from the equations of motion. We define
\beq
M = \left( \matrix{ a & b \cr
                   -b & a } \right) \equiv U + a I
\eeq
where, in this formulation, the off-diagonal matrix   $U$  appears in
the ordinary sine--Gordon system  \cite{DMH2}
while $a$ represents the noncommutative contribution.

Following closely the notation of \cite{DMH2} we introduce the chiral
matrices 
\beq
 e_{\pm} = \frac12 ( I \pm \s_3)
\eeq
so that the matrix $R$ takes 
the form $R= \sqrt{\g} e_-$. 
The linear equation
\beq
\d \chi = \l D \chi 
\label{linear2}
\eeq
with  $\chi$  an element of ${\cal M}^0$ (a  $2 \times 2$ matrix),
can be decomposed into the two equations \bea
&& (i) ~~~\pab \chi = - \l L \ast \chi \nonumber \\
&& (ii) ~~~\sqrt{\g} e_- \chi = -\l ( \pa \chi + M \ast \chi ) = 
-\l [ \pa \chi + (U + a I) \ast \chi ]
\label{system}
\eea
We look for solutions defined as power series in $\l$. As explained in 
Appendix A, $\chi$ is not invertible and therefore we will
define the currents in terms of  $\tilde{\chi} = \chi + e_-$ 
(see eq.(\ref{chitilde})). 
We introduce the matrices
\beq
j \equiv \pa \tilde{\chi} \ast \tilde{\chi}^{-1}
\qquad ; \qquad 
\tilde{j} \equiv \pab \tilde{\chi} \ast \tilde{\chi}^{-1}
\eeq

Following the details given in 
Appendix A, from the system (\ref{system}) we obtain the following
equations which have to be satisfied by $j$ and $\tilde{j}$
\bea
\sqrt{\g} j = && - \l \pa j - \l j \ast j \nonumber \\
&& + \l [ - a - e_+ U \ast a \ast U^{-1} + e_+ \pa U \ast U^{-1} 
- e_- U + e_- (\pa a) \ast U^{-1} ] \ast j \nonumber \\
&& + \l [ e_+ U \ast U - e_+ \pa a - e_+ U \ast a \ast U^{-1} \ast a 
+ e_+ \pa U \ast U^{-1} \ast a \nonumber \\
&&~~~~~~~~~~~~- e_- \pa U + e_- (\pa a) \ast U^{-1} \ast a ] 
- \sqrt{\g} e_+ a
\label{final1} 
\eea
and
\beq
\tilde{j} = - \l L \ast ( e_+ - e_- U^{-1} \ast a) + \l L \ast e_- 
U^{-1} \ast \tilde{j}
\label{final2}
\eeq
We expand $j = \sum_{l \geq 0} \l^l j^{(l)}$
and $\tilde{j} = \sum_{l \geq 0} \l^l \tilde{j}^{(l)}$. 
Substituting in the previous equations, up to the second order in $\l$
we find
\bea
&& j^{(0)} = - e_+ a \nonumber \\
&& j^{(1)} = \frac{1}{\sqrt{\g}}
e_+ U \ast U + \frac{1}{\sqrt{\g}} e_- ( U \ast a - \pa U) \nonumber \\
&& j^{(2)} =\frac{1}{\g}e_+( -U \ast \pa U +U\ast [U,a]_{\ast}) \\
&&~~~~~+\frac{1}{\g} e_-(-2 \pa U \ast a - U \ast \pa a  
+[U,a]_{\ast} \ast a -U \ast U \ast U + \pa (a \ast U) +  \pa^2 U)
\nonumber
\eea
and
\bea
&& \tilde{j}^{(0)} = 0 \nonumber \\
&& \tilde{j}^{(1)} = - L e_+ + L \ast e_- U^{-1} \ast a \nonumber \\
&& \tilde{j}^{(2)} = -L \ast e_- U^{-1} \ast L e_+ + 
L \ast e_- U^{-1} L \ast e_- U^{-1} \ast a
\eea
If we now introduce $J^{(l)} \equiv {\rm Tr} j^{(l)}$ and 
$\Tilde{J}^{(l)} \equiv {\rm Tr} \tilde{j}^{(l)}$ as the functions which
enter the definitions (\ref{recipe}) of the conserved 
currents, up to second order we find
\bea
&& J^{(0)} = -  a  \nonumber \\
&& J^{(1)} = - \frac{1}{\sqrt{\g}} \, b \ast b
\nonumber \\
&& J^{(2)} =  \frac{1}{\g} \, b\ast (\pa b - [b,a]_{\ast} )
\label{Jbig}
\eea
and 
\bea
&& \Tilde{J}^{(0)} = 0 \nonumber \\
&& \Tilde{J}^{(1)} = 
\sqrt{\g}  \, ( - \sin_{\ast}^2{\frac{\phi}{2}} -\frac12
\sin_{\ast}{\phi} \ast b^{-1} \ast a )
\nonumber \\
&& \Tilde{J}^{(2)} =\g \,
\left( \frac{1}{2}\sin_{\ast}\phi \ast b^{-1} \sin_{\ast}^2 \frac{\phi}{2}
+\frac{1}{4} \sin_{\ast}\phi \ast b^{-1} \ast   \sin_{\ast}\phi 
\ast b^{-1} \ast a \right)
\label{tildeeq}
\eea
A quite lengthy but straightforward calculation, along the same steps 
explained above, gives
\bea
J^{(3)} &=& \frac{1}{\g^{3/2}} \Big(- b \ast b \ast b \ast b + \pa b \ast 
\pa b  \nonumber \\
&~~& - \pa a \ast \pa a + \pa a \ast a \ast a + \pa a \ast b \ast b
+ \pa a \ast b^{-1} \ast \pa^2 b \nonumber \\
&~~& - b \ast a \ast \pa b - \pa b \ast b \ast a + \pa b \ast a \ast b 
+ b \ast \pa b \ast a \nonumber \\
&~~&  + 2 b \ast a \ast b \ast a - b \ast b \ast a \ast a 
- b \ast a \ast a \ast b \nonumber \\
&~~& + \pa a \ast b^{-1} \ast \pa( a \ast b) - 2 \pa a \ast b^{-1} 
\ast \pa b \ast a - \pa a \ast b^{-1} \ast a \ast b \ast a \Big)
\eea
\bea
\tilde{J}^{(3)} &=& \g^{3/2} \Big( - \frac14 \sin_{\ast}{\phi} \ast
b^{-1} \ast \sin_{\ast}{\phi} \ast b^{-1} \ast \sin^2_{\ast}{\frac{\phi}{2}}
\nonumber\\
&~~~&~~~~~~~ - \frac14 \cos^2_{\ast}{\frac{\phi}{2}} \ast b^{-1} \ast 
\sin_{\ast}{\phi} \ast b^{-1} \ast \sin_{\ast}{\phi} \nonumber \\
&~~~&~~~~~~~ -\frac{1}{8}  \sin_{\ast}{\phi} \ast b^{-1} \ast 
\sin_{\ast}{\phi} \ast b^{-1} \ast \sin_{\ast}{\phi} \ast b^{-1} \ast a
\Big)   
\eea   
We have now all the ingredients to write the first non trivial conservation 
laws associated to the equations (\ref{sg3}). Using the general recipe 
(\ref{recipe}), we find:

\noindent
1) Order zero in $\l$
\beq
{\cal J}^{(0)} = J^{(0)} = -a \quad ; \quad 
\Tilde{\cal J}^{(0)} = 0
\eeq
This current is trivially conserved ($\pab {\cal J}^{(0)} =0$)
once the equations of motion (\ref{eqmotion2}) are taken into account.

\noindent
2) Order one in $\l$ 
\beq
{\cal J}^{(1)} = J^{(1)} - \theta J^{(0)} \diamond \pa \Tilde{J}^{(1)}
 \quad ; \quad  \Tilde{\cal J}^{(1)} = \Tilde{J}^{(1)} - \theta J^{(0)}
\diamond \pab \Tilde{J}^{(1)} 
\label{conserved1}
\eeq
with $J^{(0)},J^{(1)},\Tilde{J}^{(1)}$ given in (\ref{Jbig}, \ref{tildeeq}).   
Note that, as a consequence of $\pab J^{(0)} =0$,  
the second term in $\Tilde{\cal J}^{(1)}$ is trivial 
(it is a $\pab$--derivative). 

\noindent
3) Second order in $\l$
\bea
&&{\cal J}^{(2)}= J^{(2)} - \theta (J^{(0)} \diamond \pa \tilde{J}^{(2)} +
J^{(1)} \diamond \partial \tilde{J}^{(1)}) \nonumber \\
&&\tilde{\cal J}^{(2)}= \tilde{J}^{(2)} - \theta (J^{(0)} \diamond \pab \tilde{J}^{(2)} +
J^{(1)} \diamond \pab \tilde{J}^{(1)})
\label{conserved2}
\eea

\noindent
4) Third order in $\l$
\bea
&&{\cal J}^{(3)}= J^{(3)} - \theta (J^{(0)} \diamond \pa \tilde{J}^{(3)} +
J^{(1)} \diamond \partial \tilde{J}^{(2)} + J^{(2)} \diamond \partial 
\tilde{J}^{(1)} ) \nonumber \\
&&\tilde{\cal J}^{(3)}= \tilde{J}^{(3)} - \theta (J^{(0)} \diamond \pab 
\tilde{J}^{(3)} + J^{(1)} \diamond \pab \tilde{J}^{(2)} + J^{(2)} \diamond 
\pab \tilde{J}^{(1)})
\label{conserved3}
\eea

The general argument presented  at the begining of this  Section  automatically
guarantees  that these currents are conserved.  From our procedure, they are 
obviously constructed out of our sine--Gordon field $\phi$. Furthermore, our  
construction is based on the existence of a solution $\chi$ of the linear 
system and the existence of this solution, i.e.  the bicomplex  
integrability  condition, is guaranteed only when the
field $\phi$ satisfies the system of equations (\ref{sg3}). 

More importantly, the currents ${\cal J}$, $\tilde{\cal J}$ are {\em local} 
functions of the field $\phi$.
( We use the term ``local'' in its standard meaning: the currents depend on 
the field $\phi$ and its derivatives, but not on integrals of  $\phi$.  
Of course, the intrinsic nonlocality of a noncommutative theory -- 
$\phi$--derivatives to infinite order, multiplying the parameter $\theta$ -- 
is present.) We note that the $\chi^{(l)}$ themselves are not local
in the  above sense. Indeed, it is not 
difficult to ascertain, by examining the solution of the system 
(\ref{system}),  that it will depend nonlocally on the field 
$\phi$. For example, {\em even in the commutative case}, 
the {\em trace} of the first equation integrates to
\beq
{\rm Tr} \chi (z, \bar{z}) = {\rm Tr} c(z) {\rm exp} \left[ - \l 
\int^{\bar{z}} d \bar{z}' L( \phi (z, \bar{z}'))\right]
\label{asolution}
\eeq
as can be verified by direct differentiation, and this exhibits directly the 
nonlocality we are discussing, after expanding in powers of $\l$, of  
(at least some of) the matrix elements of
$\chi^{(l)}$.  On the other hand, the quantities $j$, $\tilde{j}$  in 
(\ref{final1}, \ref{final2})
satisfy local algebraic equations and lead to local conserved currents. 
The nonlocality present in the solutions $\chi$ disappears from 
$ {\rm Tr}\bar{\pa} \chi  / \chi =
\bar{\pa} {\rm Tr} {\rm ln} \chi$ and what, superficially, appears to be a 
trivial total derivative
turns into a local function of the field $\phi$. 
(There is a subtlety here:
these equations, and their solutions, involve the $\ast$--inverse $U^{-1}$ 
and it is conceivable that 
this quantity  involves an integral; however, we must again consider this 
kind of nonlocality as acceptable since it is intrinsic to the 
noncommutativity of the theory).

\section{Perturbative expansion in $\theta$}

The currents constructed in the previous section 
 depend explicitly on the noncommutation parameter $\theta$.

To better understand their dependence on $\theta$, their connection
with the ordinary currents and the role of $b^{-1}$ in their 
expressions, we evaluate them perturbatively in $\theta$ 
and check explicitly their conservation up to second order. 
In doing this, we will make repeated use of the $\ast$--identities 
contained in Appendix B. To begin with we work out the explicit $\theta$ 
dependence of the equations of motion.

We can evaluate the expressions $a$ and $b$ by using the identity 
(\ref{identity1}). We obtain
\bea
&& a = \frac{1}{2!} \left( -\frac{i}{2} \right)^2 [ \pa \phi ,
\phi ]_{\ast} + {\rm even ~ powers}
\nonumber \\
&& b = \frac{1}{2} \pa \phi + 
\frac{i}{3!} \left( -\frac{i}{2} \right)^3 [[ \pa \phi ,
\phi ]_{\ast} , \phi ]_{\ast} + {\rm odd~powers} 
\eea
The perturbative expansion of the commutator is given in (\ref{identity5}).
Making use of that result we can write
\bea
&& a = - \frac{1}{8} 
\theta (\pa^2 \phi \, \pab \phi - \pab \pa \phi \, \pa \phi ) + O(\theta^3)
\nonumber \\
&& b = \frac12 \pa \phi 
\nonumber \\ 
&& ~~~~+ \frac{1}{48} \theta^2
\left( 2 \pab \pa^2 \phi \, \pa \phi \, \pab \phi - 
\pa^3 \phi \, (\pab \phi)^2 - \pa \pab^2 \phi \, (\pa \phi)^2
+ \pa^2 \phi \, \pab^2 \phi \, \pa \phi - (\pa \pab \phi)^2 \, \pa \phi \right)
\nonumber \\
&&~~~~~~~~~~~~~~~~~~~+ O(\theta^4)
\label{identity2}
\eea
These are formal expansions where only the dependence on $\theta$ from the 
$\ast$--product has been considered. To complete the expansion we have
to take into account also the $\theta$--dependence of the dynamical
field $\phi(z,\bar{z},\theta)$.
Writing
\beq
\phi(z,\bar{z}, \theta) = \sum_{n \geq 0} \theta^n \phi_n  
\label{phiseries}
\eeq
and inserting in the expressions (\ref{identity2}), we finally have
\bea
&& a \equiv  \sum_{n \geq 0} \theta^n a_n ~=~ - \frac{1}{8} 
\theta (\pa^2 \phi_0 \, \pab \phi_0 - \pab \pa \phi_0 \, \pa \phi_0 ) + 
\nonumber \\
&&~~~~~~~~~~~~~~~~~~- \frac{1}{8} \theta^2 ( \pa^2 \phi_0 \, \pab \phi_1 + 
\pa^2 \phi_1 \, \pab \phi_0 - \pab \pa \phi_0 \, \pa \phi_1 
- \pab \pa \phi_1 \, \pa \phi_0 ) + O(\theta^3)
\nonumber \\
&& b \equiv \sum_{n \geq 0} \theta^n b_n ~=~ 
\frac12 \pa \phi_0 + \theta \frac12 \pa \phi_1
\nonumber \\ 
&& ~~~~~~~~~~~~~~~~~+ \theta^2 \left[ \frac12 \pa \phi_2 + \frac{1}{48}
\left( 2 \pab \pa^2 \phi_0 \, \pa \phi_0 \, \pab \phi_0 - 
\pa^3 \phi_0 \, (\pab \phi_0)^2 - \pa \pab^2 \phi_0 \, (\pa \phi_0)^2
\right. \right.
\nonumber \\
&&~~~~~~~~~~~~~~~~~~~~~~~~~~~~~~
\left. \left. 
+ \pa^2 \phi_0 \, \pab^2 \phi_0 \, \pa \phi_0 - (\pa \pab \phi_0)^2 \, 
\pa \phi_0 \right) \right] + O(\theta^3)
\label{identity10}
\eea
From the equations of motion (\ref{sg3}) we can read the equations of motion
for the coefficient functions $a_n$, $b_n$. 
The $a_n$ equations are simply $\pab a_n =0$, whereas for
$b_n$ we need to use the $\theta$--expansion (\ref{sine}) for $\sin_{\ast}$
on the r.h.s. of (\ref{sg3}). Taking into account the explicit expressions
for $a_n$ and $b_n$ as read from (\ref{identity10}) we finally have that,
up to second order, the equations of motion satisfied by the various components
of $\phi$ in the $\theta$--expansion are:

\noindent
1)Order zero in $\theta$ 
\beq
a_0 = 0 ~~,~~ b_0 = \frac12 \pa \phi_0
\label{b0}
\eeq
and the equations are
\beq
\pa \pab \phi_0 = \g \sin{\phi_0}
\label{EOM0}
\eeq

\noindent
2)Order one 
\bea
&& a_1 = - \frac{1}{8} (\pa^2 \phi_0 \pab \phi_0 - \pa \pab \phi_0 \pa \phi_0)
\nonumber \\
&& b_1 = \frac12 \pa \phi_1 
\nonumber \\
&& \sin_{\ast}{\phi}|_{\theta} = \phi_1 \, \cos{\phi_0}
\label{orderone}
\eea
The equations of motion then read
\bea 
&& \pab (\pa^2 \phi_0 \pab \phi_0 - \pa \pab \phi_0 \pa \phi_0) =0
\nonumber \\
&& \pa \pab \phi_1 = \g \, \phi_1 \, \cos{\phi_0}
\label{EOM1}
\eea
We notice that, using the equation of motion at order zero, which also implies 
$\pa^2 \pab \phi_0 \, \pab \phi_0 = \pa \pab^2 \phi_0 \, \pa \phi_0$, 
the first equation in (\ref{EOM1}) can be written as
\beq
\pa^2 \phi_0 \pab^2 \phi_0 - (\pa \pab \phi_0)^2 =0
\label{identity3}
\eeq
From this identity it also follows
$[ \pa \phi_0 , \pab \phi_0 ]|_{\theta} = 0$ and $(\pa \phi_0 \ast 
\phi_0)|_{\theta^2} =0$. 

\vskip 3pt
\noindent
3)Order two: Using the previous identities for $\phi_0$ we have
\bea
&& a_2 = - \frac{1}{8} ( \pa^2 \phi_0 \, \pab \phi_1 + 
\pa^2 \phi_1 \, \pab \phi_0 - \pab \pa \phi_0 \, \pa \phi_1 
- \pab \pa \phi_1 \, \pa \phi_0 ) \nonumber \\
&& b_2 = \frac12 \pa \phi_2 + 
\frac{1}{48} \left( \pa^2 \pab \phi_0 \, \pa \phi_0 \, \pab \phi_0 -
\pa^3 \phi_0 \, (\pab \phi_0)^2 \right) 
\nonumber \\
&&~~= \frac12 \pa \phi_2 + \frac{1}{6} \pab \phi_0 \, \pa a_1 
\label{ordertwo}
\eea
and the equations of motion are
\bea
&& \pab ( \pa^2 \phi_0 \, \pab \phi_1 + 
\pa^2 \phi_1 \, \pab \phi_0 - \pab \pa \phi_0 \, \pa \phi_1 
- \pab \pa \phi_1 \, \pa \phi_0 ) =0 
\nonumber \\
&& \pa \pab \phi_2 + \frac13 \pab(\pab \phi_0 \, \pa a_1 ) = 
\g (\sin_{\ast}{\phi})|_{\theta^2}
\eea

\vskip 15pt
The conserved currents ${\cal J}^{(l)}$ are given in terms of the field 
$\phi$ and its derivatives. Therefore, using the expansion (\ref{phiseries}) 
we can write 
\beq
{\cal J}^{(l)} = \sum_{n=0}^{\infty} \theta^n {\cal J}^{(l)}_n
\label{currexp}
\eeq
where, according to the general conservation law 
(\ref{conservation}) each coefficient has to satisfy 
\beq
\pab {\cal J}^{(l)}_n = \pa \Tilde{\cal J}^{(l)}_n
\eeq
Setting $\theta=0$ in (\ref{currexp}) we should recover the ordinary conserved
currents for the commutative sine--Gordon system. Indeed, we find
${\cal J}^{(0)}_0 = 0$, while 
\beq
{\cal J}^{(1)}_0 = J_0^{(1)} = - \frac{b_0^2}{\sqrt{\g}} = 
- \frac{1}{4\sqrt{\g}} (\pa \phi_0)^2
\eeq
is the spin 2 stress tensor of the ordinary sine--Gordon system.
Its conservation law reads
\beq
\pab \left(- \frac{1}{4\sqrt{\g}} (\pa \phi_0)^2 \right) = 
\pa \left( - \sqrt{\g} \sin^2{\frac{\phi_0}{2}} \right)   
\label{J1cons}
\eeq
where the r.h.s. coincides with $\pa \Tilde{\cal J}^{(1)}_0$ as given in 
(\ref{conserved1}) for $\theta=0$. 

For the next two currents, we find ${\cal J}^{(2)}_0$ to be a total 
derivative, whereas
\beq
{\cal J}^{(3)}_0 = \frac{1}{4 \g^{3/2}} \left[ - \frac{1}{4} 
(\pa \phi_0)^4 + (\pa^2 \phi_0)^2 \right]
\eeq
It coincides with the spin 4 nontrivial current of the ordinary sine--Gordon. 
The on--shell conservation reads
\beq
\pab \left[ - \frac{1}{4} 
(\pa \phi_0)^4 + (\pa^2 \phi_0)^2 \right] = 
\pa \left[ \frac{\g}{4} (\pa \phi_0)^2 \cos{\phi_0} \right]
\eeq

\subsection{Perturbative evaluation of ${\cal J}^{(1)}$ current}

We now concentrate on the perturbative evaluation of the stress tensor
${\cal J}^{(1)}$  as given in (\ref{conserved1}).
We are interested in computing the explicit expressions of the coefficients  
${\cal J}^{(1)}_n$, $n>0$ and check their conservation
\beq
\pab {\cal J}^{(1)}_n = \pa ({\rm something})
\eeq
We will push the calculation up to second order in $\theta$.

The first non trivial deformation of the ordinary stress tensor due to 
the noncommutativity is given by
\beq
{\cal J}^{(1)}_1 = J^{(1)}_1 = -\frac{1}{\sqrt{\g}} (b \ast b)|_{\theta}
= -\frac{2}{\sqrt{\g}} b_0 b_1 =  -\frac{1}{2\sqrt{\g}} \pa \phi_0 \pa \phi_1
\eeq
Its conservation reads
\bea
\pab {\cal J}^{(1)}_1 &=& - \frac{1}{2\sqrt{\g}} \pab ( \pa \phi_0 \,
\pa \phi_1) = -\frac{1}{2\sqrt{\g}} ( \pa \pab \phi_0 \, \pa \phi_1 + 
\pa \phi_0 \, \pa \pab \phi_1 )
\nonumber \\
&=& -\frac{\sqrt{\g}}{2} ( \sin{\phi_0} \, \pa \phi_1 + \pa \phi_0 \, 
(\cos{\phi_0}) \, \phi_1 )
\nonumber \\
&=& -\frac{\sqrt{\g}}{2} \pa ((\sin{\phi_0}) \, \phi_1) 
\eea  
The last line coincides with $ \pa \Tilde{\cal J}^{(1)}_1$, up to total
derivative terms.   

At second order in $\theta$ the current is given by
\beq
{\cal J}^{(1)}_2 = J^{(1)}_2 - J^{(0)}_1 \pa \Tilde{J}^{(1)}_0
= -\frac{1}{\sqrt{\g}} (b \ast b)|_{\theta^2} - a_1 \pa( \sqrt{\g} 
\sin^2{\frac{\phi_0}{2}} )
\eeq
We evaluate $(b \ast b)|_{\theta^2}$ by observing that contributions 
at order $\theta^2$ come both from the expansion of $b$ and of the 
$\ast$--product
(see identity (\ref{identity12})). Collecting all the terms we have 
\beq
b \ast b = b_1^2 + 2 b_0 b_2 + \frac14
( \pa^2 b_0 \pab^2 b_0 - (\pa \pab b_0)^2) 
\label{J2}
\eeq
and the final expression for the stress tensor at second order is 
\beq
{\cal J}^{(1)}_2 = -\frac{1}{\sqrt{\g}} \left[ b_1^2 + 2 b_0 b_2 + \frac14
( \pa^2 b_0 \pab^2 b_0 - (\pa \pab b_0)^2)  
- \frac{\g}{2} a_1 \pa \cos{\phi_0}  \right] 
\eeq
with $a_1, b_0, b_1$ and $b_2$ given in (\ref{b0}, \ref{orderone}) and 
(\ref{ordertwo}) respectively. 

We now apply the $\pab$--derivative and prove that on--shell  
it can be written as a $\pa$--derivative of some quantity. Already at this
order the check is quite complicated but it is worth pursuing it to 
understand how noncommutativity works. First of all,
as proven in Appendix B, the following identity holds 
\beq
\pab \left[ \frac14
( \pa^2 b_0 \pab^2 b_0 - (\pa \pab b_0)^2)  
- \frac{\g}{2} a_1 \pa \cos{\phi_0}  \right] =
\pa \left[ - \frac{\g}{2} \pab \left( a_1 \cos{\phi_0} \right)\right]
\label{identity7}
\eeq
Therefore, the second and the third terms in (\ref{J2}) satisfy a
conservation equation. For the first two terms, after inserting the explicit
expressions for $b_0$, $b_1$ and $b_2$, we have
\bea
\pab ( b_1^2 + 2 b_0 b_2 ) &\equiv& \pab \left( \frac14 (\pa \phi_1)^2
+ (\pa \phi_0) \, b_2 \right) 
\nonumber \\
&=& \frac12 \pa \phi_1 \pa \pab \phi_1 + (\pa \pab \phi_0) b_2 + 
(\pa \phi_0) \pab b_2   
\nonumber \\
&=& \frac{\g}{2} (\pa \phi_1) \phi_1 \cos{\phi_0} + \frac{\g}{2} 
\pa \phi_2 \, \sin{\phi_0} + \frac{\g}{6}(\pab \phi_0) \pa a_1 \sin{\phi_0}
\nonumber \\ 
&~& + \frac{\g}{2} (\pa \phi_0) (\sin_{\ast}{\phi})|_{\theta^2} 
\label{ontheway}
\eea
where the equations of motion for $b_2$ have been used. 
We concentrate on the last term $(\pa \phi_0) (\sin_{\ast}{\phi})|_{\theta^2}$.
Since we expect it to appear in 
$(\pa \phi \ast \sin_{\ast}{\phi})|_{\theta^2}$ we first 
evaluate this expression up to second order
\bea
&& (\pa \phi \ast \sin_{\ast}{\phi})|_{\theta^2} =
\nonumber \\
&& \left( \pa \phi_0 + \theta \pa \phi_1 +\theta^2 \pa \phi_2 \right)
\ast \left( \sin{\phi_0} + \theta \phi_1 \cos{\phi_0} + \theta^2
(\sin_{\ast}{\phi})|_{\theta^2} \right)
\eea
We perform the $\ast$--product and keep only quadratic terms in $\theta$. 
Using the identity (\ref{appidentity4}) and 
\beq
\pa^3 \phi_0 \, \pab \phi_0 - \pa^2 \pab \phi_0 \, \pa \phi_0 
= - 8 \pa a_1
\eeq
which are consequences of the zero order equations of motion,
we end up with
\bea
(\pa \phi \ast \sin_{\ast}{\phi})|_{\theta^2} &=& \pa \phi_0 
(\sin_{\ast}{\phi})|_{\theta^2} - 4 a_2 \, \cos{\phi_0}  + 
4 \phi_1 a_1 \, \sin{\phi_0} 
\nonumber \\
&~& + (\pa a_1) (\pab \phi_0) \, \sin{\phi_0} + \phi_1 (\pa \phi_1) 
\cos{\phi_0} + (\pa \phi_2) \, \sin{\phi_0} 
\label{identity15}
\eea
On the other hand, from the identity (\ref{appidentity10}) proven 
in Appendix B we read
\bea
(\pa \phi \ast \sin_{\ast}{\phi})|_{\theta^2} 
&=& -\pa (\cos_{\ast}{\phi})|_{\theta^2}  + 4 \phi_1 a_1 \, \sin{\phi_0}
- \frac{2}{3} \pab (\pa a_1 \, \cos{\phi_0}) - 4 a_2 \, \cos{\phi_0}
\nonumber \\
&=& -  \pa (\cos_{\ast}{\phi})|_{\theta^2} + 4 \phi_1 a_1 \, 
\sin{\phi_0}
+ \frac{2}{3} \pa a_1 \, (\pab \phi_0)\sin{\phi_0} - 4 a_2 \, \cos{\phi_0} 
\nonumber \\
&&~~~~~~~~~
\label{identity14}
\eea
Comparing the equations (\ref{identity15}) and 
(\ref{identity14}) we finally obtain
\bea
\pa \phi_0 (\sin_{\ast}{\phi})|_{\theta^2} &=&
- \pa (\cos_{\ast}{\phi})|_{\theta^2} - 
\frac{1}{3}(\pa a_1) (\pab \phi_0) \, \sin{\phi_0}   
- \phi_1 (\pa \phi_1) \cos{\phi_0}
\nonumber \\
&~& - (\pa \phi_2) \, \sin{\phi_0} 
\eea
It is now easy to see that if we insert this result in (\ref{ontheway}) 
a lot of cancellations occur and we are left with
\bea
\pab ( b_1^2 + 2 b_0 b_2 ) = - \frac{\g}{2} \, 
\pa ( \cos_{\ast}{\phi})|_{\theta^2}
\eea
Therefore, the conservation law at second order reads
\beq
\pab {\cal J}_2^{(1)} ~=~ \pa \left( \frac{\sqrt{\g}}{2} \, 
( \cos_{\ast}{\phi})|_{\theta^2} + \frac{\sqrt{\g}}{2} \pab 
( a_1 \cos{\phi_0}) \right)
\eeq
The r.h.s. is related to $\Tilde{\cal J}_1^{(2)}$ up to total 
$\pab$--derivatives. 

To summarize, the conserved stress tensor up to second order in $\theta$
is (rescaling by a constant factor)
\bea
{\cal J}^{(1)} &=& \frac12 \pa \phi_0 \pa \phi_0 + 
\theta \pa \phi_0 \pa \phi_1 \nonumber \\
&~~& + \theta^2 \Big( \pa \phi_0 \pa \phi_2 + \frac12 (\pa \phi_1)^2
+ \frac13 \pa \phi_0 \pab \phi_0 \pa a_1 \nonumber\\
&~~&~~~~~~~~~~~~ +\frac{1}{8} \pa^3 \phi_0 \pab^2 \pa  \phi_0
-\frac{1}{8} ( \pa^2 \pab  \phi_0)^2 - \g a_1 \pa \cos{\phi_0}
\Big)
\eea
with $a_1$ given in (\ref{orderone}).

\subsection{Perturbative evaluation of ${\cal J}^{(2)}$ current}

As we have already remarked, the current 
${\cal J}^{(2)}_0$ is a total derivative, according to the well known fact
that a spin 3 current does not appear in the spectrum of the conserved
quantities for the ordinary sine--Gordon. The natural question which
arises is whether ${\cal J}^{(2)}$ remains trivial when the 
noncommutation parameter is turned on. To answer this question we compute
the first $\theta$--correction to ${\cal J}^{(2)}$.

According to the general relation (\ref{conserved2}) we have
\beq
{\cal J}^{(2)}_1 = J^{(2)}_1 - J^{(1)}_0 \pa \tilde{J}^{(1)}_0
\eeq
where, from (\ref{Jbig}) using the fact that $a$ is already order one,
\beq
J^{(2)}_1 = \frac{1}{\g} b \ast \pa b|_{\theta} = 
\pa(\frac{1}{4\g} \pa \phi_0 \pa \phi_1) + 
\frac{1}{8\g} \left( \pa^2 \phi_0 \pab \pa^2 \phi_0 - \pa \pab \phi_0
\pa^3 \phi_0 \right)
\eeq
Now, extracting $J^{(1)}_0$ and $\tilde{J}^{(1)}_0$ from eq. (\ref{J1cons}) 
we finally obtain
\beq
{\cal J}^{(2)}_1 = \pa \left( \frac{1}{4\g}\pa \phi_0 \pa \phi_1 +
\frac{1}{8} (\pa \phi_0)^2 \cos{\phi_0} - \frac{1}{8} \pa^2 \phi_0 
\sin{\phi_0} \right)
\eeq
Therefore, at first order the ${\cal J}^{(2)}$ is still trivial.  

\vskip 15pt
To summarize the results of this section, we notice that the perturbative
analysis of the conserved currents has revealed the following features:

\noindent
a) The spin of the conserved currents are the same as
the ordinary ones. Therefore, the spin spectrum of the corresponding 
integrals of motion 
\beq
Q^{(s)} = \int dz {\cal J}^{(s)} + \int d\bar{z} \Tilde{\cal J}^{(s)}
\eeq
still coincides with the exponents of the $SU(2)$
algebra, modulo the Coxeter number.  
This means that noncommutativity does not affect the algebraic structure
which underlies the model. 

\noindent
b) Despite the appearance of $b^{-1}$ in the general expression for
$\Tilde{\cal J}^{(1)}$, the conservation law at spin 2, order by order in
$\theta$, only involves the $\ast$--product of $\phi$ and its derivatives,
but not their inverses.

\section{Localized solutions}

Since the system of equations (\ref{sg3}) describes a constrained field, 
 the class of solutions for $\phi$ will be in general smaller than
the one corresponding to the unconstrained case. In order to show that,
in spite of the constraint, nontrivial solutions exist, 
we look for localized solutions of the equations of motion up to first 
order in $\theta$, $\phi = \phi_0 + \theta \phi_1$,
\bea
&& \pa \pab \phi_0 = \g \sin{\phi_0} 
\nonumber \\
&& \pa \pab \phi_1 = \g \phi_1 \, \cos{\phi_0} 
\qquad \qquad \pa^2 \phi_0 \pab ^2 \phi_0 - (\pa \pab  \phi_0)^2 = 0
\label{soliton0}
\eea 
Although the bicomplex approach  has given us suitable 
equations of motion which guarantee integrability, we have not been able yet 
to find the corresponding action from which they can be derived 
(but see below).
Consequently, we  use the term ``localized'' as follows: we construct the 
standard euclidean solitons at order zero in $\theta$. Since the solution 
at order zero determines the solutions at succeeding
orders, we  call ``localized'' the all-orders solution constructed in this 
fashion.  

The last equation in (\ref{soliton0}) is automatically satisfied by {\em any} 
function of
$(x^1-ivx^0) = \frac{i}{\sqrt{2}} [ -z ( 1+v) + \bar{z} (1-v)]$ 
\footnote{It is easy to show that also at second order in $\theta$
the equation of motion which does not contain the potential (first equation
in (\ref{ordertwo})) is automatically satisfied by any function of 
$(x^1-ivx^0)$.
We  conjecture that this pattern   repeats at every order.}.   
In particular, it is automatically satisfied by any function of $x^1$ only. 
To solve the other equations we first reduce the problem to a 
one--dimensional problem
by looking for solutions of equations of motion which do not depend on $x^0$. 
We then need to solve
\beq
\phi''_0 = 2\g \, \sin{\phi_0}
\qquad \qquad
\phi''_1 = 2\g \phi_1 \, \cos{\phi_0} 
\eeq
The first one is the ordinary equation for static euclidean solitons of the 
sine--Gordon system. Thus, we
apply the standard procedure to integrate it. A first integration gives
\beq
\phi'_0 = \pm 2 \sqrt{2\g} \, \sin{\frac{\phi_0}{2}}
\label{soliton1}
\eeq
A second integration (taking the plus sign in the previous equation) gives 
the one--(anti)soliton solutions
\bea
&& \phi_0^{sol}(x^1) = 4 \, {\rm arctg} \, {e^{\sqrt{2\g}(x^1 - \bar{x}^1)}} 
\nonumber \\
&& \phi_0^{antisol} (x^1) = -4 \, {\rm arctg} \, {e^{\sqrt{2\g}(x^1 - 
\bar{x}^1)}}
\label{soliton2}
\eea
We now look for solutions to the second equation in (\ref{soliton0}). 
Taking the product 
\beq
\phi'_1 \phi''_0 + \phi'_0 \phi''_1 = (\phi'_0 \phi'_1 )'
\eeq
and inserting the equations of motion on the l.h.s. we can perform a first 
integration to obtain
\beq
\phi'_0 \phi'_1 = 2 \g \phi_1 \sin{\phi_0}
\label{soliton6} 
\eeq
Inserting eq. (\ref{soliton1}) on the l.h.s. and dividing by $\phi_1$, 
we  have
\bea
\frac{\phi'_1}{\phi_1} &=&  \sqrt{2 \g}  \cos{\frac{\phi_0}{2}} 
\nonumber \\
&=&  \sqrt{2 \g} \,
\frac{1 - e^{2\sqrt{2\g}(x^1 - \bar{x}^1)}}{1 + 
e^{2\sqrt{2\g}(x^1 - \bar{x}^1)}}
\eea
This equation can be easily integrated and gives 
\beq
\phi_1(x^1) = \frac{1}{\ch{\sqrt{2\g}(x^1 - \bar{x}^1)}}
\label{soliton3}
\eeq
A plot of this function for $2\g =1$, $\bar{x}^1=0$ is given in Fig. 1.

\vskip 18pt
\noindent
\begin{minipage}{\textwidth}
\begin{center}
\includegraphics[width=0.60\textwidth]{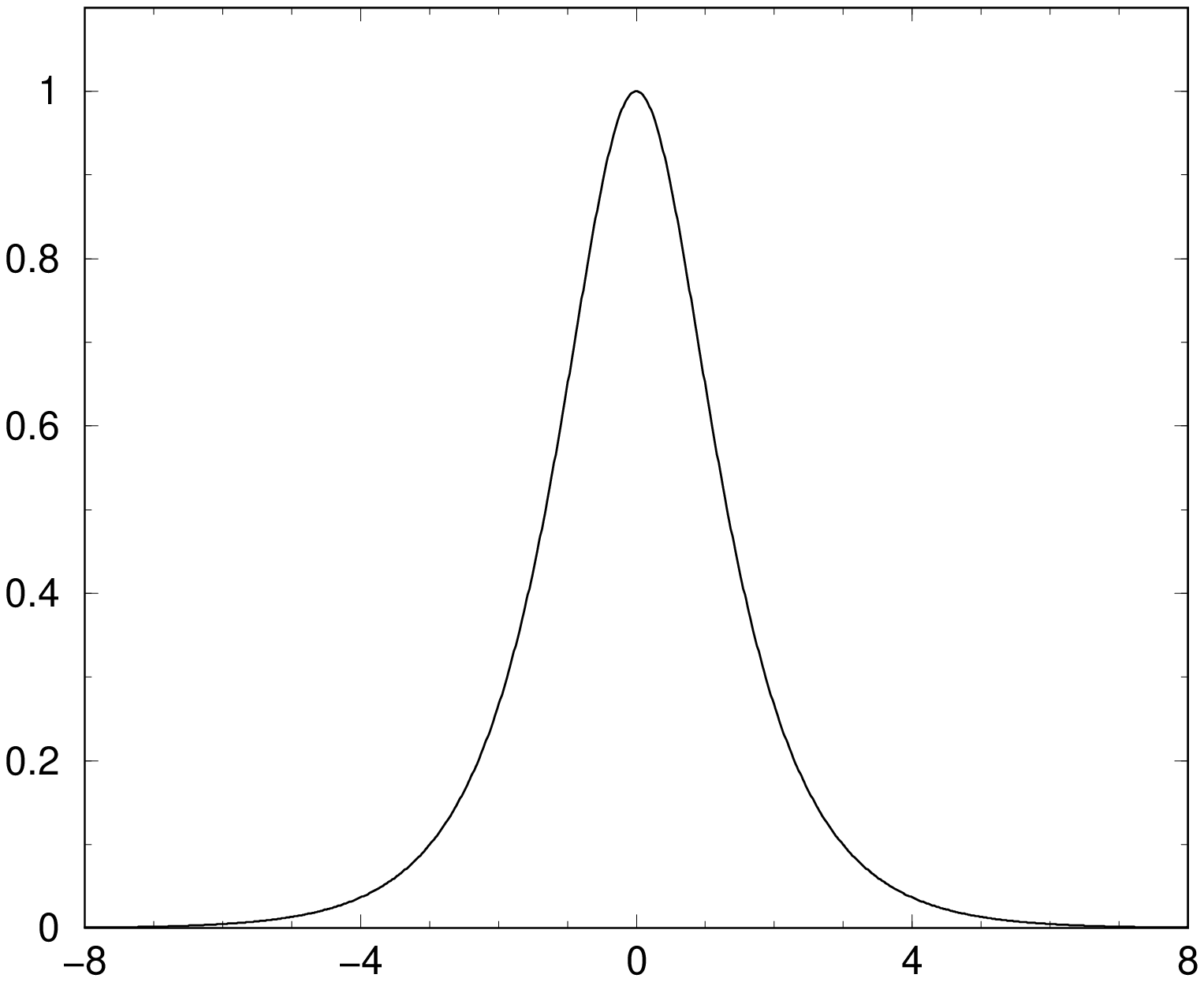}
\end{center}
\begin{center}
{\small{Figure 1:
First order correction to the one--(anti)soliton solution}}
\end{center}
\end{minipage}

\vskip 20pt

It represents the first order correction generated by  noncommutativity
to the euclidean ``one--soliton'' solutions of the ordinary sine--Gordon 
equation. The first order correction to the antisoliton solution is
again (\ref{soliton3}). 

In Fig. 2 we have plotted the ``one--soliton'' solution (again $2\g=1$ and 
$\bar{x}^1=0$) at first order in $\theta$
for different values of the deformation parameter .
As it can be easily seen, the perturbation due to the noncommutativity
mainly affects the ``soliton'' around $x^1=0$, while leaving  its asymptotic
behavior at large $|x^1|$ unmodified. 

This fact has immediate consequences for the topological charge. 
Even for noncommutative solitons we can define the topological charge as
\beq
{\cal T} = \frac{1}{2\pi} \int_{-\infty}^{+\infty} dx^1 \frac{d \phi}{dx^1}
\equiv \sum_n \theta^n {\cal T}^{(n)} 
\eeq 
where $\phi$ is the localized solution at all orders in $\theta$.
For $\theta=0$ the topological charge is ${\cal T}^{(0)}=1$ for the soliton 
and ${\cal T}^{(0)}=-1$ for the antisoliton. Computing it for the first order 
correction (\ref{soliton3}) we find ${\cal T}^{(1)} =0$.  At this order
noncommutativity does not affect the topological properties of the
solution.

It is interesting to note that the first two equations of motion in 
(\ref{soliton0}) can be derived from the following action
\beq
S = \int d^2x \, \left\{ \frac12 \pa \phi_0 \pab \phi_0 + 
\g ( 1 -\cos{\phi_0}) + \theta [ \pa \phi_0 \pab \phi_1 + \g \phi_1 
\sin{\phi_0} ] \right\} 
\eeq
For static solutions, using (\ref{soliton1}, \ref{soliton6}) we can write 
\beq
S = 2 \g \, \int dx^0 dx^1 \, \left[ 
(1 - \cos{\phi_0}) ~+~ \theta \, \phi_1 \sin{\phi_0} \right]
\eeq

\vskip 18pt
\noindent
\begin{minipage}{\textwidth}
\begin{center}
\includegraphics[width=0.60\textwidth]{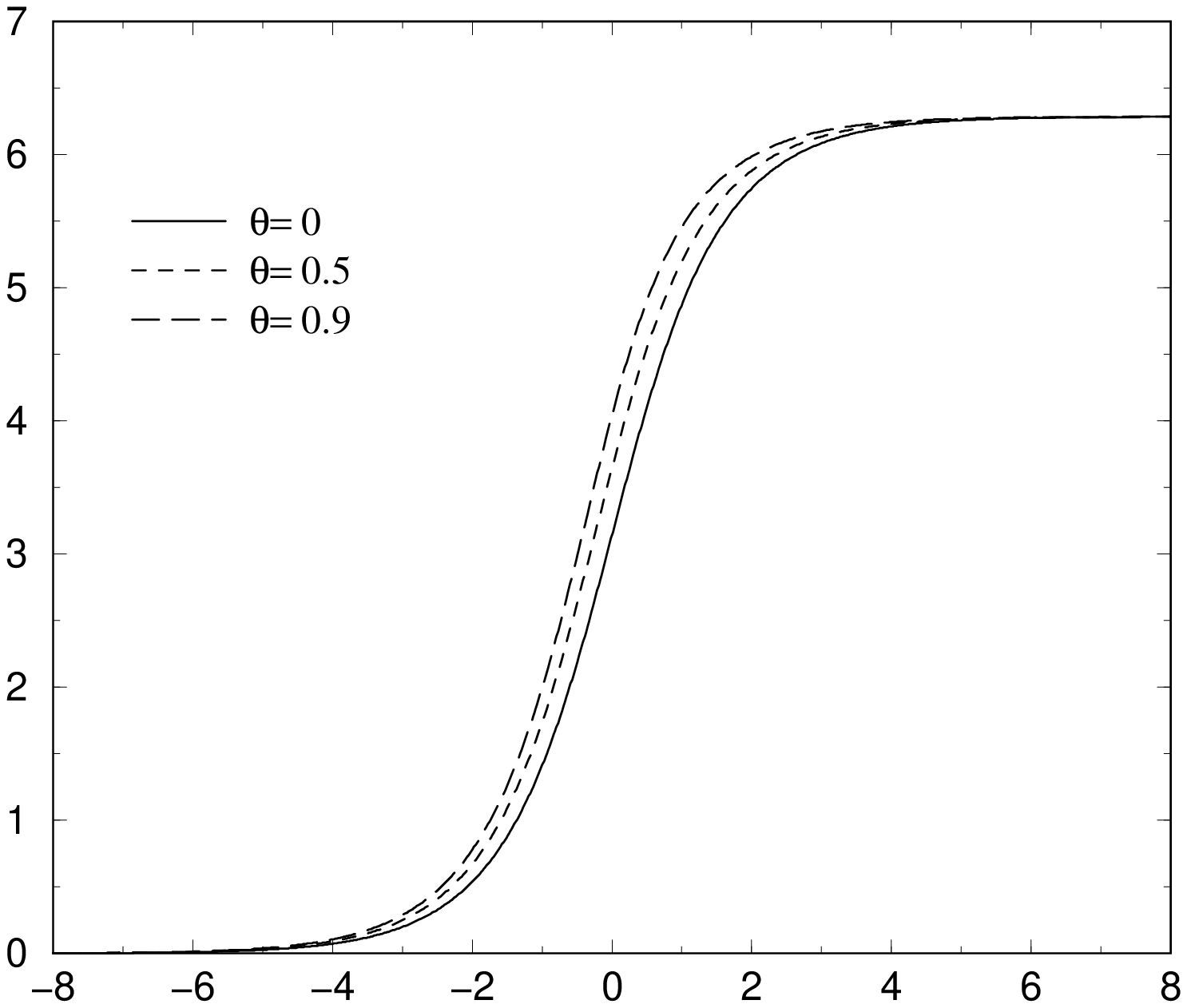}
\end{center}
\begin{center}
{\small{Figure 2:
One--soliton solution including the first order correction for 
various values of $\theta$}}
\end{center}
\end{minipage}

\vskip 20pt

\noindent
If we compactify the system on a cylinder $0 < x^0 < 2\pi$, $-\infty 
< x^1 <\infty$, for the soliton (\ref{soliton2}, \ref{soliton3}) we obtain
\beq
S = \sqrt{2\g}\int_0^{2\pi} dx^0 \left[ -4 \frac{1}{1 + e^{2\sqrt{2\g}
(x^1 - \bar{x}^1)}} ~+~ \theta \frac{1}{\ch^2{\sqrt{2\g}(x^1-\bar{x}^1)}}
\right] \Big|_{-\infty}^{+\infty}~=~ 8\pi \sqrt{2\g} 
\eeq
which is the value of the action for the ordinary sine--Gordon
euclidean soliton. We then conclude that at this order noncommutativity does
not change the value of the classical action. It would be interesting to
investigate whether this is a peculiarity of the first order or it is a 
general feature. 

Starting from the solutions (\ref{soliton2}, \ref{soliton3}) which do not 
depend 
on the euclidean time $x^0$, we can generate ``nonstatic'' solutions. As 
already noticed, the 
third eq. in (\ref{soliton0}) constrains $\phi_0$ to depend on $(x^1-ivx^0)$,
so that 
\bea
&& \phi_0^{sol} (x^0, x^1) = 4 \, 
{\rm arctg} \, {e^{\sqrt{2\g}\frac{(x^1 - \bar{x}^1
-iv x^0)}{\sqrt{1-v^2}}}} 
\nonumber \\
&& \phi_0^{antisol}(x^0, x^1) = -4 
\, {\rm arctg} \, {e^{\sqrt{2\g}\frac{(x^1 - \bar{x}^1
-iv x^0)}{\sqrt{1-v^2}}}} 
\label{soliton4}
\eea
is easily verified to be a solution of the first eq. in 
(\ref{soliton0}). For the $x^0$ dependence of $\phi_1$, at this order we do not
have any constraint. However, since $\phi_0$ enters its equation of motion, 
we expect $\phi_1$ to have the same dependence on $x^0$. Indeed,
by direct inspection, one  realizes that a ``nonstatic'' solution is
\beq
\phi_1(x^0, x^1) = \frac{1}{\ch{(\sqrt{2\g}\frac{(x^1 - \bar{x}^1
-iv x^0)}{\sqrt{1-v^2}}})}
\label{soliton5}
\eeq  

Finally, we  look for a generalization of Backlund transformations to the 
noncommutative case. In the ordinary case, these are first order equations 
which generate multi--``soliton'' solutions starting from a given solution
with fewer ``solitons''. 

Consider first the case $\theta=0$. Localized solutions 
$\Tilde{\phi}_0$ of the equations of motion (\ref{soliton0}) can be generated 
by solving the first order equations
\bea
&& \frac12 \pa (\Tilde{\phi}_0 - \phi_0 ) = \a \, \sqrt{\g} \, \sin{
\frac{(\Tilde{\phi}_0 + \phi_0 )}{2}} 
\nonumber \\
&& \frac12 \pab (\Tilde{\phi}_0 + \phi_0 ) = \frac{1}{\a} \, \sqrt{\g} \, 
\sin{\frac{(\Tilde{\phi}_0 - \phi_0 )}{2}}
\label{backlund0}
\eea
where $\phi_0$ is a known solution and $\a$ an arbitrary real parameter.
Indeed, by applying $\pab$ to 
the first equation it is easy to check that $\Tilde{\phi}_0$ satisfies
the equations of motion, once $\phi_0$ does. 

At the first order in $\theta$, given a solution $\phi_1$, we look
for a function $\Tilde{\phi}_1$ which satisfies 
\bea
&& \frac12 \pa (\Tilde{\phi}_1 - \phi_1 ) = \a \, \sqrt{\g} \, 
\frac{\Tilde{\phi}_1 + \phi_1}{2} \, \cos{\frac{(\Tilde{\phi}_0+\phi_0 )}{2}} 
\nonumber \\
&& \frac12 \pab (\Tilde{\phi}_1 + \phi_1 ) = \frac{1}{\a} \, \sqrt{\g} \, 
\frac{\Tilde{\phi}_1 - \phi_1}{2} \, \cos{\frac{(\Tilde{\phi}_0-\phi_0 )}{2}} 
\label{backlund1}
\eea
where $\tilde{\phi}_0$ satisfies (\ref{backlund0}).
Again, by application of $\pab$ to the first equation, it is easy to verify
that $\Tilde{\phi}_1$ is a solution of the equation of motion at first order. 
This system can be used to generate first order corrections to 
multi--``soliton'' solutions.

\section{Conclusions and outlook} 

We have discussed in this paper  an integrable noncommutative two-dimensional field 
theory whose equations of motion reduce to the ordinary sine-Gordon equation
in the commutative limit.
In considering generalizations of the ordinary sine-Gordon system 
to the NC case one might be tempted to start from the  usual action
$\int d^2x[ \frac{1}{2} \pa \phi \pab \phi + \g(1- \cos{\phi} )]$ and 
replace ordinary products by $\ast$--products \cite{NOS}.
However, since
the currents obtained as a natural extension of the ordinary ones are
not conserved, the corresponding system is not guaranteed to be integrable.
Instead our approach,  which expresses the field equations as integrability 
conditions of a   bicomplex system, constructs directly  classically 
conserved currents which reduce to the standard currents of the commutative 
theory in the limit of $\theta \to 0$.

Constructing a NC extension of sine--Gordon directly at the level of
equations of motion, necessarily generates a NC system of two equations,
one of them being a natural extension of the ordinary one, whereas the other,
which has the structure of a constraint equation, has only a NC origin. 
The appearance of the second equation is quite unavoidable and seems to be
necessary in order to guarantee integrability. It is a 
consenquence of the fact that, in the NC case, the $SU(2)$ symmetry group 
of ordinary sine--Gordon is enlarged to $U(2)$ which contains a noncommutative
$U(1)$ factor. Therefore, the group valued fields involved in eq. (\ref{eq})
have a nontrivial trace part which is responsible for the appearance
of the constraint equation. 
  
We have recursively constructed an infinite set of
conservation laws. In the present approach, writing down a conservation 
equation 
for suitably 
defined objects is not a difficult task. What is essential however is that 
these objects be {\em local}, i.e. not involve integrals of the 
field $\phi$.  Our currents, defined in terms of {\em nonlocal} solutions 
of a certain bidifferential equation, satisfy this requirement.
We have performed some  calculations to verify the conservation of  the 
currents to low order in the parameter $\theta$. In particular, from our 
results it appears that noncommutativity does not affect the spin spectrum
of the conserved currents. 

We have presented  ``localized'' solutions, to first order  in the NC 
parameter, as well as the corresponding Backlund transformations. For 
the former, a better understanding of their significance and of the 
corresponding topological charges would require a knowledge of the 
classical action. We were able to make some progress by working to 
first order in $\theta$.

Since the system we are describing is a constrained system,
the class of solutions for the field $\phi$ will be in general smaller 
than the one of the
corresponding unconstrained system (the system which would satisfy only 
the second 
equation in (\ref{sg3})). This is already clear at the perturbative level 
where for instance, the constraint at first order (see last equation in 
(\ref{soliton0}))
selects a particular subclass among all the solutions $\phi_0$ of the ordinary 
sine--Gordon equation (even if  we have shown that it does not restrict 
the spectrum of 
localized solutions). We are faced  with
 the question  whether our NC system can be considered a natural NC 
extension of the sine--Gordon system.  
To answer this question, we should define what we mean in general with 
``NC extension''
of a field theory. The most natural definition is that the NC system 
should reproduce
the ordinary field theory when the limit $\theta \to 0$ is done appropriately. 
However, our example shows that there can be situations where the limit is 
not smooth.
In fact, if we take the limit at the level of equations of motion, we obtain
the ordinary equations and, consequently, the whole spectrum of ordinary 
solutions.   
If we perform the limit directly on the solutions we seem to lose
part of the ordinary spectrum. A similar pattern can be observed in the NC
extension of the $U(1)$ WZNW model \cite{WZNW, NOS}, since the NC 
equations of motion contain
a nontrivial constraint which restricts the set of solutions. Also in 
this case,
performing the $\theta \to 0$ limit at the level of solutions one does not 
recover the whole dynamics of the ordinary model.    

The main motivation of our work was to construct a NC system which is 
integrable and is related to the commutative sine-Gordon system. 
Therefore we concentrated on the currents and the equations of 
motion which guarantee their conservation. As we mentioned above, 
a direct NC generalization of the sine-Gordon action appears not 
to lead to a set of conserved currents.
At the same time, it is not trivial to find an action from which 
our equations of motion follow. Although we were able to do this 
to first order in the $\theta$ parameter, this issue remains an 
open problem. Furthermore, even if such an action were found, 
or guessed at (and we do have some possible candidates), because 
manipulations involving NC exponentials are so cumbersome, checking 
that it would lead to the correct equations of motion might be an 
equally difficult task.

A final comment: 
It is evident, from  what  we have presented in this paper, that explicit 
calculations in NC theories are rather complicated. It is to be hoped that
better techniques -- a $\ast$--calculus -- can be developed to facilitate 
manipulations
which are trivial in the commutative case but extremely difficult here. 
This seems essential if one has any hope to move on to a quantum formulation 
of the theory.

\medskip

\section*{Acknowledgements}
\noindent 
We would like to thank the referee, whose objections and comments 
were quite helpful in focusing our thoughts on some of the issues.

\noindent This work has been supported in 
part by INFN, MURST and the European Commission RTN program
HPRN--CT--2000--00131, in which S.P. is associated to
the University of Padova. M.T.G acknowledges support by the NSF 
under grant PHY-00-70475.

\newpage

\appendix
\section{Derivation of conserved currents}

In this Appendix we give the detailed derivation of the equations
(\ref{final1}, \ref{final2}) which have been used in the text to obtain the
conserved currents. 

We concentrate on the system (\ref{system}) which, for convenience, 
we write again
\bea
&& (i) ~~~\pab \chi = - \l L \ast \chi \nonumber \\
&& (ii) ~~~\sqrt{\g} e_- \chi = -\l ( \pa \chi + M \ast \chi ) = 
-\l [ \pa \chi + (U + a I) \ast \chi ]
\label{systemapp}
\eea
Setting $\l=0$, from
eq. $(ii)$ we see that at zero order $\chi$ satisfies $e_- \chi^{(0)} =0$. 
A solution is then $\chi^{(0)}=e_+$. For $\l \neq 0$ instead, we
apply $e_+$ to eq. $(ii)$  
to obtain (in the derivation we make often use of
the identity $e_+ U = U e_-$)
\beq
e_+ \pa \chi = - e_+ (U + a I) \ast \chi = - U \ast (e_- \chi) - 
e_+ a \ast \chi 
\label{eminus}
\eeq
Substituting $e_- \chi$ as given by eq. $(ii)$ we can write
\beq
\sqrt{\g} e_+ \pa \chi = 
\l [ U \ast \pa \chi + U \ast U \ast \chi +U \ast a \ast \chi ]
-\sqrt{\g} a \ast e_+ \chi    
\eeq
This can be added to $\sqrt{\g} e_- \pa \chi$ obtained by differentiating
eq. $(ii)$ 
\beq
\sqrt{\g} e_- \pa \chi = - \l [ \pa^2 \chi + \pa U \ast \chi + U \ast \pa \chi
+ \pa a \ast \chi + a \ast \pa \chi ] 
\eeq
to write
\bea
\sqrt{\g} \pa \chi & = &\l [ - \pa^2 \chi - a \ast \pa \chi + U \ast U 
\ast \chi + 
U \ast a \ast \chi - \pa U \ast \chi - \pa a \ast \chi ] \nonumber \\
&~&~~~~ -\sqrt{\g} a \ast e_+ \chi  
\label{chieq}
\eea
Since $\chi^{(0)} = e_+$ is not invertible, it follows that $\chi$ will not 
be invertible for  $\l \to 0$. Therefore, we consider instead the shifted 
function
\beq
\tilde{\chi} = \chi + e_-
\label{chitilde}
\eeq
Computing $e_-\chi$ from eq. (\ref{eminus}) we have 
\beq
\chi = e_+ \chi + e_- \chi = 
e_+ \tilde{\chi} - e_- U^{-1} \ast \pa \tilde{\chi} - e_- U^{-1} \ast a
\ast \tilde{\chi}
\label{chiminus}
\eeq
Substituting in (\ref{chieq}) we find a differential equation
for $\tilde{\chi}$.
If we multiply that equation by $\tilde{\chi}^{-1}$ we obtain an 
equation 
for $j \equiv \pa \tilde{\chi} \ast \tilde{\chi}^{-1}$. Making use of the  
identity $\pa^2 \tilde{\chi} \ast \tilde{\chi}^{-1} = \pa j + j \ast j$
we finally have
\bea
\sqrt{\g} j = && - \l \pa j - \l j \ast j \nonumber \\
&& + \l [ - a - e_+ U \ast a \ast U^{-1} + e_+ \pa U \ast U^{-1} 
- e_- U + e_- (\pa a) \ast U^{-1} ] \ast j \nonumber \\
&& + \l [ e_+ U \ast U - e_+ \pa a - e_+ U \ast a \ast U^{-1} \ast a 
+ e_+ \pa U \ast U^{-1} \ast a \nonumber \\
&&~~~~~~~~~~~~- e_- \pa U + e_- (\pa a) \ast U^{-1} \ast a ] 
- \sqrt{\g} e_+ a
\label{apfinal1} 
\eea
which is eq. (\ref{final1}) in the text. 

We now go back to the system (\ref{systemapp}) and consider eq. $(i)$. 
Using (\ref{chiminus})  we find
\beq
\pab \tilde{\chi} = - \l L \ast ( e_+ \tilde{\chi} - e_- U^{-1} \ast
\pa \tilde{\chi} - e_- U^{-1} \ast a \ast \tilde{\chi} )
\eeq
and for $\tilde{j} \equiv \pab \tilde{\chi} \ast \tilde{\chi}^{-1}$   
\beq
\tilde{j} = - \l L \ast ( e_+ - e_- U^{-1} \ast a) + \l L \ast e_- 
U^{-1} \ast \tilde{j}
\label{apfinal2}
\eeq
which is eq. (\ref{final2}) used in the text.

\section{$\ast$--calculus}

In this Appendix we collect and prove  identities which are
useful when dealing with the NC equations and their 
$\theta$--expansion.

We start by recalling that $\ast$--functions are defined in terms of their
$\ast$--series expansion. In particular, we have
\bea
&& e_{\ast}^{a \phi} = \sum_{n=0}^{\infty} \frac{a^n}{n!} \phi \ast
\phi \ast \cdots \ast \phi \equiv 
\sum_{n=0}^{\infty} \frac{a^n}{n!} \phi^n_\ast 
\nonumber \\
&& \cos_{\ast}{\phi} = \frac{e_{\ast}^{i\phi} + e_{\ast}^{-i \phi}}{2}
\nonumber \\
&& \sin_{\ast}{\phi} = \frac{e_{\ast}^{i\phi}  - e_{\ast}^{-i \phi}}{2i}
\label{A3}
\eea
As a consequence of the general identity
\beq
e_{\ast}^{a \phi} \ast e_{\ast}^{b \phi} =  e_{\ast}^{(a+b) \phi} 
\eeq
which follows from the definition of the $\ast$--exponential,
it is easy to prove that the  main trigonometric identities are still
valid. Among them we list
\beq
\cos_{\ast}^2{\phi} + \sin_{\ast}^2{\phi} = 1
\eeq
\bea
\cos_{\ast}^2{\frac{\phi}{2}} = \frac{1 + \cos_{\ast}{\phi}}{2}
\nonumber \\
\sin_{\ast}^2{\frac{\phi}{2}} = \frac{1 - \cos_{\ast}{\phi}}{2}
\eea
\beq
\sin_{\ast}{\frac{\phi}{2}} \ast \cos_{\ast}{\frac{\phi}{2}} =
\cos_{\ast}{\frac{\phi}{2}} \ast \sin_{\ast}{\frac{\phi}{2}} =
\frac12 \sin_{\ast}{\phi}
\label{A1}
\eeq
On the other hand, due to the lack of commutativity, the derivatives of 
exponentials and trigonometric functions do not satisfy the
nice properties they have in the commutative case. 
This is due to the fact that the derivative of the exponential is not 
proportional to the exponential itself; instead\beq
\pa e_{\ast}^{a \phi} = a \int_0^1 dt e_{\ast}^{t a\phi} \ast
\pa \phi \ast  e_{\ast}^{(1-t)a \phi} 
\label{A2}   
\eeq
It follows that $\pa \cos_{\ast}{\phi} \neq -\sin_{\ast}{\phi}$ and
the check of the conservation laws at any order in $\theta$ is a hard   
problem.

We now list the main identities which can be useful to write the explicit
expressions of the currents perturbatively in $\theta$. Since in the
text we perform perturbative calculations up to second order, we will stop our
identities at that order. We give formal expansions in terms of the field
$\phi$ and its derivatives, keeping in mind that $\phi$ itself may 
depend on $\theta$ and eventually 
must be expanded in power series in the noncommutation parameter.
We have
\beq
\phi \ast \phi = \phi^2 + \frac{\theta^2}{4} \left( \pa^2 \phi \pab^2 \phi 
- (\pa \pab \phi)^2 \right) + O(\theta^3) 
\label{identity12} 
\eeq
\beq
[ \pa \phi , \phi ]_{\ast} = \theta ( \pa^2 \phi \diamond \pab \phi - 
\pab \pa \phi \diamond \pa \phi ) 
=\theta (\pa^2 \phi \, \pab \phi - \pab \pa \phi \, \pa \phi ) + O(\theta^3)
\label{identity5}
\eeq
where the $\diamond$--product has been defined in (\ref{diamond}).
Other useful identities are
\beq
e_{\ast}^{\frac{i}{2} \phi} \ast \pa e_{\ast}^{-\frac{i}{2} \phi} =
- \frac{i}{2} \pa \phi + \frac{1}{2!} \left( -\frac{i}{2} \right)^2 
[ \pa \phi ,
\phi ]_{\ast} + \frac{1}{3!} \left( -\frac{i}{2} \right)^3 [[ \pa \phi ,
\phi ]_{\ast} , \phi ]_{\ast} + \cdots
\label{identity1}
\eeq
\bea
\sin_{\ast}{\phi} &=& \sin{\phi} + O(\theta^2) 
\nonumber \\
&=& \sin{(\phi_0 + \theta \phi_1 + \cdots)}  + O(\theta^2) 
\nonumber \\
&=&  \sin{\phi_0} + \theta \phi_1 \, \cos{\phi_0}  + O(\theta^2)
\label{sine}
\eea
\bea
\sin^2_{\ast}{\frac{\phi}{2}} &=& \frac{1 - 
\cos_{\ast}{\phi}}{2} 
\nonumber \\
&=& \frac12 - \frac12 \left( \cos{\phi_0} - \theta \, \phi_1 \, \sin{\phi_0}  
\right) + O(\theta^2)
\label{cosine}
\eea
Less trivial identities which have been used in checking the conservation 
of ${\cal J}_1$ at second order are (\ref{identity7}) and (\ref{identity14}).
Here we give a proof of these two identities. 

A way to check (\ref{identity7}) is to compute 
$(\pa \phi_0 \ast \sin{\phi_0})|_{\theta^2}$ , where $\sin$ is the ordinary 
sine, in two different ways:

\noindent
1) using the definition of star product at that order
\bea
&& (\pa \phi_0 \ast \sin{\phi_0})|_{\theta^2}  
\nonumber \\
&& = \frac{1}{8} ( \pa^3 \phi_0 \, \pab^2 \sin{\phi_0} + \pab^2 \pa \phi_0 
\, \pa^2 \sin{\phi_0} - 2 \pa^2 \pab \phi_0 \, \pa \pab \sin{\phi_0} ) 
\nonumber \\
&& = - \frac{1}{8} ( \pa^3 \phi_0 \, \pab \phi_0 - \pab \pa^2 \phi_0 \, \pa 
\phi_0)
\, \pab \phi_0 \, \sin{\phi_0}
= \pa a_1 \, \pab \phi_0 \, \sin{\phi_0}
\nonumber \\
&&= - \pab ( \pa a_1 \cos{\phi_0})
\eea   
Note that the equations of motion at order zero and one have been used 
and also the following identity 
\beq
\pa^3 \phi_0 \, \pab^2 \phi_0 + \pab^2 \pa \phi_0 \, \pa^2 \phi_0
- 2\pa^2 \pab \phi_0 \, \pa \pab \phi_0 = \pa ( \pa^2 \phi_0 \pab^2 \phi_0 -
(\pa \pab \phi_0)^2) = 0
\label{appidentity4}
\eeq
which follows from (\ref{identity3}).

\noindent
2) using the equations of motion from the very beginning
\bea
&& (\pa \phi_0 \ast \sin{\phi_0})|_{\theta^2}  
\nonumber \\
&& = \frac{1}{\g} ( \pa \phi_0 \ast \pa \pab \phi_0)|_{\theta^2}
= \frac{1}{2\g} \pab( \pa \phi_0 \ast \pa \phi_0)|_{\theta^2}
\nonumber \\
&& = \frac{1}{8\g}
\pab ( \pa^3 \phi_0 \, \pab^2 \pa \phi_0 - (\pab \pa^2 \phi_0)^2 ) 
\eea
Comparing the results from the two procedures we obtain
\beq
\pab ( \pa^3 \phi_0 \, \pab^2 \pa \phi_0 - (\pab \pa^2 \phi_0)^2 ) 
= - 8\g \pab ( \pa a_1 \cos{\phi_0})
\label{appidentity1}
\eeq
which is the identity (\ref{identity7}).

To check the identity (\ref{identity14})
we evaluate $\pa(\cos_{\ast}{\phi})|_{\theta^2} = 
(\pa \cos_{\ast}{\phi})|_{\theta^2}$. By expanding the cosine in power series,
we are left with the evaluation of   
\beq
\pa(\phi \ast \cdots \ast \phi)|_{\theta^2}
= \left\{ (\pa \phi) \ast \cdots \ast \phi + 
\phi \ast (\pa \phi) \ast \cdots \ast \phi
+ \cdots \phi \ast \cdots \ast (\pa \phi) \right\}_{\theta^2}
\eeq
for the $\ast$--product of $2n$ fields. It can be written as
\beq
\pa(\phi \ast \cdots \ast \phi)|_{\theta^2}
= \left\{ 2n (\pa \phi) \ast \phi_{\ast}^{2n-1}  + \sum_{j=1}^{2n-1} 
[ \phi_{\ast}^j , \pa \phi ]_{\ast} \ast \phi_{\ast}^{2n-1-j} 
\right\}_{\theta^2}
\label{appidentity2}
\eeq
At this order we can replace any $\ast$--power $\phi_{\ast}^m$ in the
sum with the ordinary power $\phi^m$ (note that the commutator 
is already order one in
$\theta$ and $\phi_{\ast}^m = \phi^m + O(\theta^2)$ according to 
(\ref{identity12})).
Now consider the following identity
\bea
&& [ \phi^j , \pa \phi ]_{\ast}|_{{\rm up~to~}\theta^2} =
\nonumber \\
&& \theta j (\phi_0 + \theta \phi_1)^{j-1} \left\{ (\pa \phi_0 + \theta
\pa \phi_1)(\pa \pab \phi_0 + \theta \pa \pab \phi_1) \right.
\nonumber \\
&&~~~~~~~~~~~~~~~~~~~~~~~~~~ \left.
- (\pab \phi_0 + \theta \pab \phi_1)(\pa^2 \phi_0 + \theta \pa^2 \phi_1)
\right\}|_{{\rm up~to~}\theta^2}
\nonumber \\
&&= 
\theta \, 8j \, \phi_0^{j-1} \, a_1 
+ \theta^2 \left( 8j (j-1) \, \phi_0^{j-2} \, \phi_1 \, a_1 + 8j \, 
\phi_0^{j-1} \, a_2 \right)
\eea
We substitute in (\ref{appidentity2}), perform the $\ast$--product and keep 
only $\theta^2$--order terms
\bea
&& \pa(\phi \ast \cdots \ast \phi)|_{\theta^2}
= 2n (\pa \phi) \ast \phi_{\ast}^{2n-1}|_{\theta^2}  
\nonumber \\
&~& + \sum_{j=1}^{2n-1} 
\left[ 8j(j-1) \phi_0^{2n-3} \, \phi_1 \, a_1 + 8j(2n-1-j) \phi_0^{2n-3} \, 
\phi_1 \, a_1 + 8j \phi_0^{2n-2} \, a_2 \right] 
\nonumber \\
&~& +  \sum_{j=1}^{2n-2} 8j \frac12 \left[ \pa( \phi_0^{j-1} \, a_1) 
\pab \phi_0^{2n-1-j} - \pab ( \phi_0^{j-1} \, a_1) \pa \phi_0^{2n-1-j} 
\right] 
\eea
In the last term, using the equations of motion $\pab a_1 =0$, 
we are left with 
$(2n-1-j) \phi_0^{2n-3} \pa a_1 \pab \phi_0$. 
Now perform the sums over $j$ and substitute in the original equation
to find
\bea
&&(\pa \cos_{\ast}{\phi})|_{\theta^2} =
\nonumber \\
&~& \sum_{n=1}^{\infty} \frac{(-1)^n}{(2n)!} \, 2n (\pa \phi) \ast 
\phi_{\ast}^{2n-1} \Big|_{\theta^2}
~+~ 8 \sum_{n=2}^{\infty} \frac{(-1)^n}{(2n)!} \, n (2n-1)(2n-2)
\phi_0^{2n-3} \, \phi_1 \, a_1 
\nonumber \\
&+& \frac{4}{3} \sum_{n=2}^{\infty} \frac{(-1)^n}{(2n)!} \, n (2n-1)
(\pab \phi_0^{2n-2}) \, \pa a_1 
~+~  8 \sum_{n=1}^{\infty} \frac{(-1)^n}{(2n)!} \, n (2n-1) \phi_0^{2n-2}
\, a_2
\nonumber \\
&=& - (\pa \phi \ast \sin_{\ast}{\phi})|_{\theta^2} 
+ 4 \phi_1 a_1 \, \sin{\phi_0} - \frac{2}{3} \pab 
(\pa a_1 \, \cos{\phi_0}) - 4 a_2 \, \cos{\phi_0}
\label{appidentity10}
\eea
as claimed.

\end{document}